\pdfoutput=1
\documentclass[10pt,journal,a4paper]{IEEEtran}
\usepackage[english]{babel}
\usepackage{fixltx2e-fixes}
\usepackage[LGR, T1]{fontenc}
\usepackage{ucs}
\usepackage[utf8x]{inputenc}
\usepackage{textcomp}
\usepackage[artemisia]{textgreek}
\usepackage[T1]{url}
\usepackage[unicode,pdftex]{hyperref}
\usepackage[final]{microtype}
\usepackage{amsmath}
\usepackage[thmmarks, amsmath]{ntheorem}
\usepackage[romanConstants, boldVectors]{scstuff}
\usepackage{units}
\usepackage{xspace}
\usepackage{mdwtab}
\usepackage{acronym}
\usepackage{cite}
\usepackage[pdftex]{graphicx}
\usepackage[svgnames]{xcolor}
\usepackage{paralist}
\usepackage[caption=false]{subfig}
\usepackage{amsmath-matrix-ext}

\hypersetup{colorlinks=true, 
  linkcolor=DarkBlue,
  citecolor=DarkBlue,
  urlcolor=DarkBlue}
\newcommand\email[1]{\href{mailto:#1}{\nolinkurl{#1}}}

\graphicspath{{./}{../Figures/}{./Figures/}}

\makeatletter
\newcommand\isacroused[3]{%
  \expandafter\ifx\csname ac@#1\endcsname\AC@used
  #2\else #3\fi}
\newcommand\setsubcapskip[1]{\sf@capskip #1}
\makeatother
\newacro{FIR}{Finite Impulse Response}
\newacro{IIR}{Infinite Impulse Response}
\newacro{STF}{Signal Transfer Function}
\newacro{NTF}{Noise Transfer Function}
\newacro{LP}{Low-Pass}
\newacro{BP}{Band-Pass}
\newacro{KYP}{Kalman–Yakubovich–Popov}
\newacro{A/D}{analog to digital}
\newacro{D/D}{digital to digital}
\newacro{D/A}{digital to analog}
\newacro{DDSM}[DΔΣM]{Digital ΔΣ Modulator}
\newacro{PLL}{Phase Locked Loop}
\newacro{CMQ}{Classical Model of Quantization}
\newacro{PSD}{Power Spectral Density}
\newacroplural{PSD}{Power Spectral Densities}
\newacro{OSR}{Oversampling Ratio}
\newacro{DTFT}{Discrete Time Fourier Transform}
\newacro{IDTFT}{\isacroused{DTFT}{Inverse DTFT}{%
    Inverse Discrete Time Fourier Transform}}

\newcommand{\NTF}{\ensuremath{\mathit{NTF}}}
\newcommand{\STF}{\ensuremath{\mathit{STF}}}
\newcommand{\FF}{\ensuremath{\mathit{FF}}}
\newcommand{\FB}{\ensuremath{\mathit{FB}}}

\newcommand\OSR{\ensuremath{\mathit{OSR}}}

\theorembodyfont{\itshape}
\theoremheaderfont{\normalfont\bfseries}
\theoremseparator{}

\theoremstyle{nonumberplain}
\theoremheaderfont{\normalfont\bfseries}
\theorembodyfont{\normalfont}
\theoremsymbol{~~\IEEEQEDopen}

\PrerenderUnicode{©}

\begin{document}
\title{Noise Weighting in the Design of \texorpdfstring{ΔΣ}{Delta-Sigma} 
  Modulators\\ (with a Psychoacoustic Coder as an Example)}

\author{%
  \thanks{This is a post-print version of a paper appearing in the IEEE
    Transaction on Circuits and Systems - Part II. Available through DOI
    \url{http://dx.doi.org/10.1109/TCSII.2013.2281892}. Always cite as the
    published version.\vspace*{0.5ex}}%
  \thanks{Copyright © 2013 IEEE. Personal use of this material is permitted.
    However, permission to use this material for any other purposes must be
    obtained from the IEEE by sending an email to
    \protect\email{pubs-permissions@ieee.org}.\vspace*{0.5ex}}%
  Sergio~Callegari,\extrainfo{~\IEEEmembership{Senior~Member,~IEEE}} and %
  Federico~Bizzarri,\extrainfo{~\IEEEmembership{Member,~IEEE}}%
  \thanks{%
    S.~Callegari is with the Advanced Research Center on Electronic Systems for
    Information and Communication Technologies ``E. De Castro'' (ARCES) at the
    University of Bologna, Italy. E-mail:
    \protect\email{sergio.callegari@unibo.it}.}%
  \thanks{%
    F.~Bizzarri is with the Dipartimento di Elettronica, Informazione e
    Bioingegneria at Politecnico di Milano, Italy. E-mail:
    \protect\email{federico.bizzarri@polimi.it}.}%
  \afterauthortext }

\initializeplaintitle[%
  \def\Titlesize{}
  \def\texorpdfstring#1#2{#2}]
\hypersetup{pdftitle=\plaintitle, 
  pdfauthor={Sergio Callegari, Federico Bizzarri}}

\markboth{IEEE Transactions on Circuits and Systems II}%
{Callegari \MakeLowercase{\textit{et al.}}: %
  Noise Weighting in the Design of ΔΣ Modulators\dots}
%



\IEEEaftertitletext{\vspace*{-0.2\baselineskip}}
\def\afterauthortext{\vspace*{-0.9\baselineskip}}

\maketitle

\begin{abstract}
  A design flow for ΔΣ modulators is illustrated, allowing quantization noise
  to be shaped according to an arbitrary weighting profile. Being based on
  \acs{FIR} \acsp{NTF}, possibly with high order, the flow is best suited for
  digital architectures. The work builds on a recent proposal where the
  modulator is matched to the reconstruction filter, showing that this type of
  optimization can benefit a wide range of applications where noise (including
  in-band noise) is known to have a different impact at different
  frequencies. The design of a multiband modulator, a modulator avoiding DC
  noise, and an audio modulator capable of distributing quantization artifacts
  according to a psychoacoustic model are discussed as examples. A software
  toolbox is provided as a general design aid and to replicate the proposed
  results.
\end{abstract}
\acresetall
\acused{SNR}
\acused{rms}
\acused{ac}
\acused{A/D}
\acused{D/A}
\acused{D/D}

\section{Introduction}
\label{sec:intro}
\begin{figure}[t]
  \begin{tightcenter}
    \includegraphics[scale=0.73]{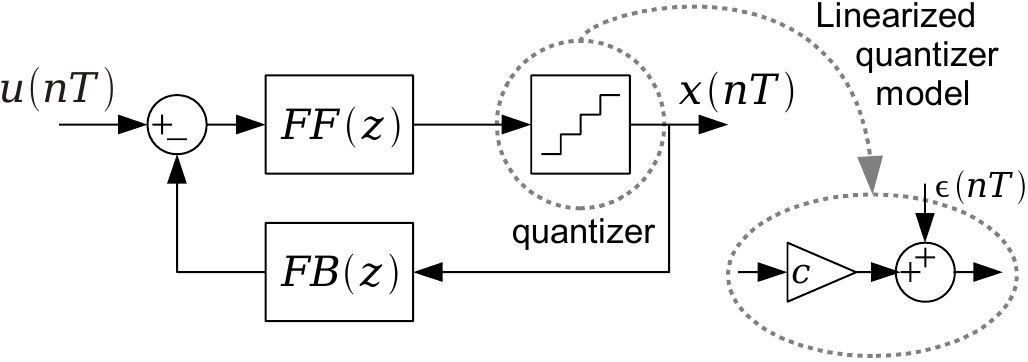}
  \end{tightcenter}
  \caption{General block diagram of a discrete-time ΔΣ modulator and
    approximated linear model.}
  \label{fig:ds-block}
\end{figure}

While often associated to \ac{A/D} conversion, ΔΣ modulators are in fact coders
that enjoy a wide range of applications. Generally speaking, they exploit a
nonlinear feedback architecture (Fig.~\ref{fig:ds-block}) to translate an
analog or high-resolution digital input into a low-resolution high-sample-rate
digital signal with minimal loss of fidelity \cite{Harris:WET-2003}. As a
matter of fact, their prevalent commercial deployment is as \emph{digital
  units}, namely \acp{DDSM} \cite{Pamarti:TCAS1-54-3}, used in tasks such as
\ac{D/A} and \ac{D/D} conversion \cite{Harris:WET-2003}, fractional \acp{PLL}
\cite{Su:TCAS2-56-12}, etc.

A key phase in their design is the selection of the \ac{NTF}
\cite{Norsworthy:DSDC-1996}, fundamental to the preservation of information
content. This is particularly true for \acp{DDSM}, whose digital nature frees
the designer from many limitations. However, the Literature is mostly concerned
with analog architectures and many of its \ac{NTF} considerations are over
constrained or not directly applicable to \acp{DDSM} \cite{Pamarti:TCAS1-54-3}.

Recall that the \ac{NTF} derives from a linear approximation replacing the
modulator quantizer with the superposition of a noise signal ($\epsilon(nT)$ in
Fig.~\ref{fig:ds-block}, where $T$ is the sample period). Together with the
\ac{CMQ} \cite{Schreier:UDSDC-2004}, this lets the modulator behavior be
expressed by two items: the \ac{STF}, from input $u(nT)$ to output $x(nT)$, and
the \ac{NTF}, from $\epsilon(nT)$ to $x(nT)$. In principle, full preservation
of information is possible if $u(nT)$, passed through the \ac{STF}, is
decoupled in band (i.e., separable by a linear filter) from $\epsilon(nT)$
passed through the \ac{NTF}. In practice, a full decoupling is never
possible. Even if it were, an \emph{ideal} filter to separate away the
quantization noise would not be realistically available
\cite{Callegari:TCAS1-60-9}. Thus, the designer should select the \emph{best
  possible} (\emph{good enough}) \ac{NTF} given actual conditions.

In this paper, a design strategy for the \ac{NTF} is proposed allowing the
quantization noise (including the residual in-band noise) to be shaped
according to an arbitrary weighting (\emph{cost}) profile.  This is not
possible with conventional flows \cite{Schreier:UDSDC-2004, Kenney:AICSP-3,
  Schreier:DELSIG}, which merely distinguish between the \emph{signal band} and
\emph{out of band frequencies}, aiming just at concentrating noise in the
latter. While in principle valid for analog modulators too, the proposed
strategy is best suited for \acp{DDSM} since it delivers a \ac{FIR} \ac{NTF},
which may easily require a high order.  The work builds on
\cite{Callegari:TCAS1-60-9} that, owing to the interpretation of \acp{DDSM} as
heuristic optimizers for \emph{filtered-approximation} problems
\cite{Callegari:TSP-58-12, Bizzarri:ISCAS-2010}, suggests that modulators can
be matched to their output/reconstruction filters (as exemplified by the
applications in \cite{Callegari:ICECS-2012, Bizzarri:ISCAS-2012}). Here, we
introduce a change in perspective illustrating how the same type of
optimization can also benefit a wide range of applications where an {\em
  explicit, tangible} filter does not exist but nevertheless there are known
reasons for suffering more from some types of noise than from others. To do so,
we emphasize by an alternative but equivalent mathematical derivation that even
when a filter can be identified only its magnitude response determines the
\ac{NTF} design, so that the response can be reinterpreted as a weighting.

The applicability of the proposed design flow is wide: (i) using \emph{on-off}
weighting functions (1 in the signal band and 0 out of it) it reproduces
conventional design methods; (ii) it deals with situations that cannot be
easily managed otherwise, such as \emph{multiband signals}; (iii) most
important, it simplifies the management of the \emph{sub-ideal behaviors} that
characterize real world applications, letting the residual in-band noise be
concentrated where it can be less harmful and the out-of-band noise be
concentrated where the removal can be more efficient.

The paper provides many application examples including the design of a coder
for audio applications distributing the residual in-band noise according to a
standard psychoacoustic model \cite{ISO226-2003}.  Directions are also provided
to download open source code meant for the replication of the results in the
examples and as a general aid for the \ac{DDSM} designer.

\section{Background}
\label{sec:background}
When a modulator such as that in Fig.~\ref{fig:ds-block} is linearized, the
relationships between $\NTF(z)$, $\STF(z)$ and the loop filters $\FF(z)$,
$\FB(z)$ are
\begin{equation}
  \begin{cases}
    \NTF(z)=\frac{1}{1+c\FF(z)\FB(z)}\\
    \STF(z)=\frac{c \FF(z)}{1+c\FF(z)\FB(z)}
  \end{cases}\hspace{-1ex}
  \begin{cases}
    \FF(z)=\frac{\STF(z)}{c\NTF(z)}\\
    \FB(z)=\frac{1-\NTF(z)}{\STF(z)}
  \end{cases}
\end{equation}
where $c=1$ is customarily assumed. This means that the \ac{NTF} selection
fully determines the modulator whenever the \ac{STF} is
pre-assigned. Typically, specifications want $\STF(z)$ to be unitary or at most
an integer delay $z^{-d}$ with $d\in\Nset{N}$. In the following, $\STF(z)$
shall be assumed to be $1$ for simplicity.

\ac{CMQ} states that uniform quantization can be approximately modeled as the
superimposition of noise, white in spectrum, independent from the quantized
signal and uniformly distributed within
$[\nicefrac{-\Delta}{2},\nicefrac{+\Delta}{2}]$, where $\Delta$ is the
quantization step. It holds relatively well whenever the modulator input signal
is ``busy'' \cite{Schreier:UDSDC-2004}. With this, the input noise power is
$\sigma^2_\epsilon=\nicefrac{\Delta^2}{12}$, and its \ac{PSD} is uniform and
equal to $\Psi_\epsilon(f)=\nicefrac{\Delta^2}{6}$, with $f$ normalized in
$[0,\nicefrac{1}{2}]$. Consequently, the noise component at the modulator
output has a \ac{PSD}
\begin{equation}
  \Psi_{n}(f) = \frac{\Delta^2}{6} \abs{\NTF\left(\ee^{\ii 2\pi f}\right)}^2 .
\end{equation}

The \ac{NTF} choice is subject to some constraints. First of all, the modulator
loop cannot be algebraic. Thus, $c \FF(z)\FB(z) =
\nicefrac{(1-\NTF(z))}{\NTF(z)}$ must include some delay. This condition can
only be satisfied if the \ac{NTF} impulse response has a unitary zero lag
coefficient. Secondly, one must guarantee that the modulator loop is
stable. This condition is hard to tackle since it is not sufficient to look at
the approximated linear model. However, it is known that a frequent mechanism
that can break the modulator operation is the \emph{overloading} of the
quantizer \cite{Schreier:UDSDC-2004}. Thus, a common approach to favor
stability consists in limiting the peak gain of the \ac{NTF}, taking
\begin{equation}
  \norm{\NTF}_\infty = \max_{f\in[0,\frac{1}{2}]} \abs{\NTF\left(\ee^{\ii 2\pi
        f}\right)} < \gamma
  \label{eq:Lee}
\end{equation}
where $\gamma$ is a constant set from the quantizer resolution. Binary
quantizers need $\gamma\leq 2$ with $1.5$ being often used. This condition,
known as \emph{Lee criterion} \cite{Lee:Thesis-1987} is neither necessary nor
sufficient for correct operation, but is empirically known to work well in a
large variety of practical cases.

\section{Noise weighting in the \ac{NTF} selection}
\label{sec:theory}
Conventional design flows assume that the \ac{NTF} can be selected from the
input signal properties only. They take the signal band $\set B$ as a starting
point to deliver an \ac{NTF} such that the quantization noise is attenuated as
much as possible in $\set B$, and thus pushed to
$[0,\nicefrac{1}{2}]\setminus\set B$, where this complement set is sufficiently
large thanks to the \ac{OSR}.
All these design flows neglect two aspects:
\begin{asparaenum}[(i)]
\item no matter how hard they try, they will never succeed in removing
  \emph{all} the quantization noise from the signal band, since this would
  require a brick-wall \ac{NTF}. Even if the latter could be obtained, no
  real-world modulator would fully obey to it;
\item even if the whole of the quantization noise could be pushed out of the
  signal band, it would be impossible to ignore it, since no real world filter
  would be able to fully remove it.
\end{asparaenum}
Furthermore, conventional strategies typically assume that $\set B$ is an
interval and cannot be easily extended to multi-band cases.

One can immediately see that the way in which the residual in band quantization
noise distributes can matter and that the way in which the noise pushed out of
band gets distributed is important as well. Actually, these considerations have
been clear for a long time. Already in 1997, \cite{Dunn:JAES-45-4} attempted at
designing ΔΣ modulators for audio applications capable to distribute the
residual in-band quantization noise so that it could be minimally
audible. However, the refinement of these older attempts is somehow
limited. More recently, \cite{Nagahara:TSP-60-6} proposed a formal method to
minimize the peak values of the in-band residual noise (namely to minimize the
maxima of $\abs{\NTF\left(\ee^{\ii 2\pi f}\right)}$). Furthermore,
\cite{Callegari:TCAS1-60-9} proposed an \emph{output filter aware} design
strategy where the modulator \ac{NTF} is matched to the filter in charge of
removing the out-of-band noise.

Here, we propose the introduction of a cost factor $\nu$ based on a weighting
function $w:[0,\nicefrac{1}{2}]\to\Nset{R}^+$
\begin{equation}
  \nu = \int_0^{\frac{1}{2}} \Psi_{n}(f) w(f)\, df
  \label{eq:merit-factor}
\end{equation}
so that the \ac{NTF} design can be based on its minimization. This lets points
(i) and (ii) be flexibly managed. Furthermore, it obviously represents a
generalization of conventional design strategies as well as specialized ones
like \cite{Dunn:JAES-45-4}, since $w(f)$ can be taken on-off or shaped
according to given profiles such as equal-loudness ones \cite{ISO226-2003,
  wannamaker:JAES-40-7}.  With respect to \cite{Callegari:TCAS1-60-9} it can be
viewed as change in perspective since the design strategy becomes aware of a
cost function, which may not have a corresponding block in the physical
architecture of the system.

To deal with the minimization, the \ac{NTF} needs to be restricted to a
\ac{FIR} form, to guarantee that non-convex expressions, which would be
unmanageable, are avoided \cite{Callegari:TCAS1-60-9}. A $P$ order \ac{NTF}
(leading to a $P$ order modulator) is described by $P+1$ coefficients $a_0,
\dots, a_{P}$ as in
\begin{equation}
  \NTF(z) = \sum_{i=0}^P a_i z^{-i} \ .
  \label{eq:ntfz}
\end{equation}
This form also guarantees the causality of the modulator filters
\cite{Callegari:TCAS1-60-9}. From \eqref{eq:ntfz}, the \ac{NTF} frequency
response is immediately obtained as the \ac{DTFT} of the coefficients
\cite{Oppenheim:DTSP-1989} like $\NTF\left(\ee^{\ii 2\pi f}\right) =
\sum_{i=0}^P a_i\, \ee^{\ii 2\pi f i}$.  Thus,
{\allowdisplaybreaks
  \begin{multline}
    \nu = \frac{\Delta^2}{6} \int_0^{\frac{1}{2}} 
    \abs{\sum_{i=0}^P a_i\, \ee^{\ii 2\pi f i}}^2  w(f)\, df =\\
    \frac{\Delta^2}{6} \sum_{i=0}^P \sum_{j=0}^P a_i a_j
    \int_0^{\frac{1}{2}} w(f)\, \ee^{\ii 2\pi f (i-j)}\, df.
  \end{multline}}
The latter can be concisely re-expressed in matrix form as $\nu =
\frac{\Delta^2}{6}\, \vec a\transposed \mat {\hat Q}\, \vec a$, where the
superscript $\transposed$ indicates transposition. To do so it is sufficient to
let $\vec a = (a_0, \dots, a_P)\transposed$ and $\mat{\hat Q}=(\hat q_{i,j})$
be a $(P+1)\times(P+1)$ matrix with entries
\begin{equation}
  \hat q_{i,j} = \int_0^{\frac{1}{2}} w(f)\, \ee^{\ii 2\pi f (i-j)}\, df .
\end{equation}
Since $\hat q_{i,j}$ only depends on $i-j$, $\mat{\hat Q}$ is
Toeplitz. Furthermore, $\hat q_{j,i} = \hat q^*_{i,j}$ where the asterisk
indicates conjugation, so that $\mat{\hat Q}$ is Hermitian. With this, $\vec
a\transposed \mat{\hat Q}^*\, \vec a = \vec a\transposed \mat {\hat Q}\, \vec a
= \nicefrac{1}{2}\; \vec a\transposed \mat {Q}\, \vec a$, with $\mat
{Q}=\mat{\hat Q}+ \mat{\hat Q}^*$, where $\mat{\hat Q}^*$ is the adjoint of
$\mat{\hat Q}$.  Thus
\begin{equation}
  \nu = (\nicefrac{\Delta^2}{3})\, \vec a\transposed \mat {Q}\, \vec a
\end{equation}
where $\mat Q$ is a Toepliz symmetric, positive defined real matrix, fully
determined by its first row, whose entries are $q_{0,j} = 2 \Re\left(
  \int_0^{\frac{1}{2}} w(f)\, \ee^{-\ii 2\pi f j}\, df \right)$.
Interestingly, by extending $w(f)$ for negative $f$ values, so that
$w(-f)=w(f)$, one also has
\begin{equation}
  q_{0,j} = \int_{-\frac{1}{2}}^{\frac{1}{2}} w(f)\, \ee^{-\ii 2\pi f j}\, df .
\label{eq:first_row}
\end{equation}
from which $\mat Q$ eventually appears to be defined by the first $P+1$ entries
of the \ac{IDTFT} of $w(f)$.

Note that the cost factor $\nu$ herein derived is equivalent to that used in
\cite{Callegari:TCAS1-60-9}. In fact, such paper assumes that an output filter
is identifiable after the modulator and aims at matching the modulator design
to its features expressed via the impulse response $h_0, h_1, h_2, \dots$. Yet,
one can quickly see that matching is optimized via the minimization of a
quadratic form based on a matrix whose entries are
$q_{j,k}=\sum_{i=-\infty}^{\infty} h_{i-j} h_{i-k}$
\cite[Eqn.~(17)]{Callegari:TCAS1-60-9}. These can be easily recognized as
self-correlation entries of the impulse response. The discrete form of the
Wiener-Khinchin theorem \cite{Oppenheim:DTSP-1989} lets this self-correlation
be related to the \ac{PSD} of the impulse response itself. Thus, in
\cite{Callegari:TCAS1-60-9} the matching could have been eventually based on
the magnitude response of the filter. Here, we change the perspective by
re-interpreting the squared magnitude response as an abstract weighting
function, to disengage from application frameworks where a tangible output
filter is recognizable. The current derivation of $\mat Q$ emphasizes this
possibility by doing away with a filter-based notation altogether. Furthermore,
it quickly provides an operative, accurate way to get $\mat Q$ from $w(f)$ via
the \ac{IDTFT} operator. Incidentally, note that any attempt at getting $\mat
Q$ from the Fast Fourier Transform of samples of $w(f)$ would be prone to
artifacts from aliasing.

Once the goal function is established, the optimal \ac{NTF} coefficients can be
determined as in \cite{Callegari:TCAS1-60-9, Nagahara:TSP-60-6}. After $a_0$ is
set to $1$, the Lee criterion is managed by defining matrices $\mat A$, $\mat
B$, $\mat C$ and $\mat D$ such that: $\mat A$ is $P\times P$ with entries
$a_{i,j}=\delta_{i,j+1}$ (where $\delta$ is the Kronecker delta); $\mat B$ is a
column vector with $P$ entries, all null but the last one set to $1$; $\mat C$
is $(a_P, a_{P-1}, \dots, a_1)\transposed$ and $\mat D=(a_0)$. With this, the
\ac{KYP} lemma \cite{Iwasaki:TAC-50-1}, lets the criterion be re-expressed as
the existence of a square symmetric positive defined matrix $\mat P$ such that
\begin{equation}
  \renewcommand\arraystretch{0}
  \begin{pmatrix}
    \mat A\transposed\,\mat P\,\mat A - \mat P &
    \mat A\transposed\,\mat P\,\mat B &
    \mat C\transposed\\
    \mat B\transposed\,\mat P\,\mat A &
    \mat B\transposed\,\mat P\,\mat B - \gamma^2&
    \mat D\\
    \mat C & \mat D & -1
  \end{pmatrix} \le 0
\end{equation} 
where `$\le$' is used to state negative semi-definiteness. Altogether, one has
a semidefinte program \cite{Boyd:LMISCT-1994} that can be tackled by interior
point methods \cite{Andersen:OML-3-2011}.

\section{Examples}
\label{sec:examples}
\subsection{Replication of conventional design flows}
\label{ssec:on-off}
\begin{figure}[b]
  \begin{tightcenter}\vskip -2ex%
    \setsubcapskip{-1ex}%
    \subfloat[\label{sfig:on-off}]{%
      \includegraphics[scale=0.75]{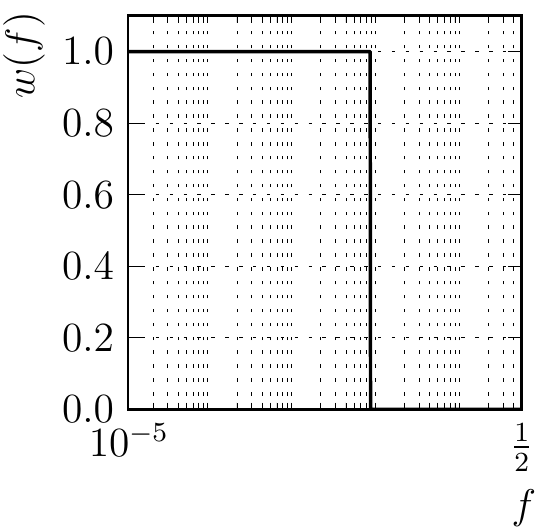}}\hfill
    \subfloat[\label{sfig:on-off-ntf}]{%
      \includegraphics[scale=0.75]{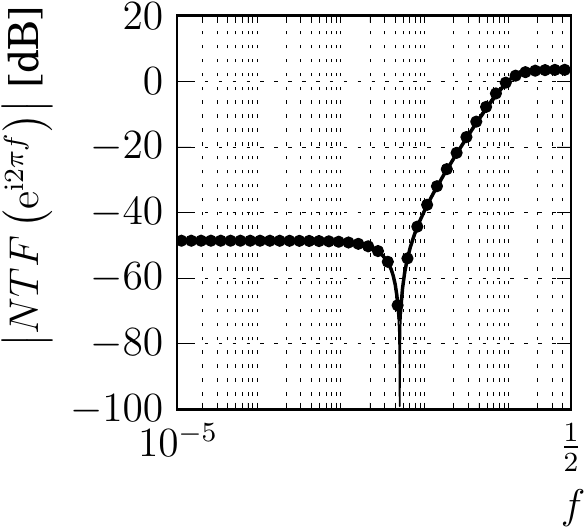}}    
  \end{tightcenter}
  \caption{Sample on-off specification for an \ac{LP} ΔΣ modulator with
    $\OSR=64$ \protect\subref{sfig:on-off} and corresponding \ac{NTF} magnitude
    response \protect\subref{sfig:on-off-ntf}. In
    \protect\subref{sfig:on-off-ntf}, the solid line is obtained with the
    proposed method for a 10\Us{th} order \ac{FIR} \ac{NTF}, while the dots are
    obtained with DELSIG's \texttt{synthesizeNTF} for a 2\Us{nd} order \ac{IIR}
    \ac{NTF}.}
  \label{fig:on-off}
\end{figure}

Conventional design flows typically take as input a modulator type, e.g.,
\ac{LP} or \ac{BP}, and an \ac{OSR}. These specification can be trivially
converted into on-off weightings. For instance, consider the design of an
\ac{LP} modulator with $\OSR=64$. This straightforwardly converts into $w(f)$
equal to $1$ for $f<\nicefrac{1}{128}$ and $0$ otherwise, as in
Fig~\ref{sfig:on-off}. Fig.~\ref{sfig:on-off-ntf} shows the magnitude response
of a 10\Us{th} order \ac{FIR} \ac{NTF} obtained by the proposed approach, and
compares it to that of a 2\Us{nd} order \ac{NTF} obtained by DELSIG's
\texttt{synthesizeNTF} \cite{Schreier:UDSDC-2004}. Obviously, for the \ac{FIR}
based approach the order must be higher. However, the two plots show that the
ability to replicate the results of conventional design flows is almost
perfect.

\subsection{Design of a multiband modulator}
\begin{figure}
  \begin{tightcenter}\vskip -2ex%
    \setsubcapskip{-1ex}%
    \subfloat[\label{sfig:on-off-multi}]{%
      \includegraphics[scale=0.75]{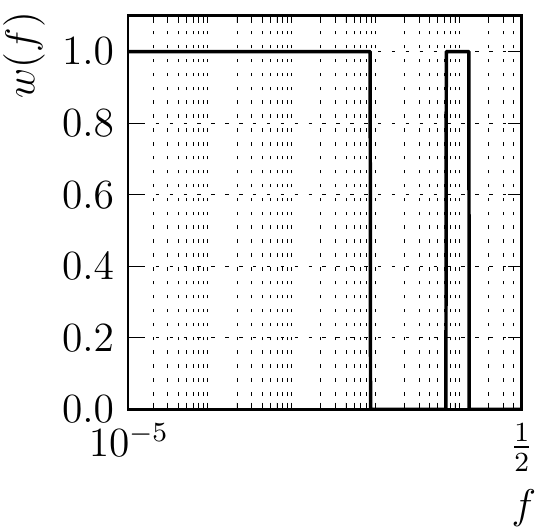}}\hfill
    \subfloat[\label{sfig:on-off-multi-ntf}]{%
      \includegraphics[scale=0.75]{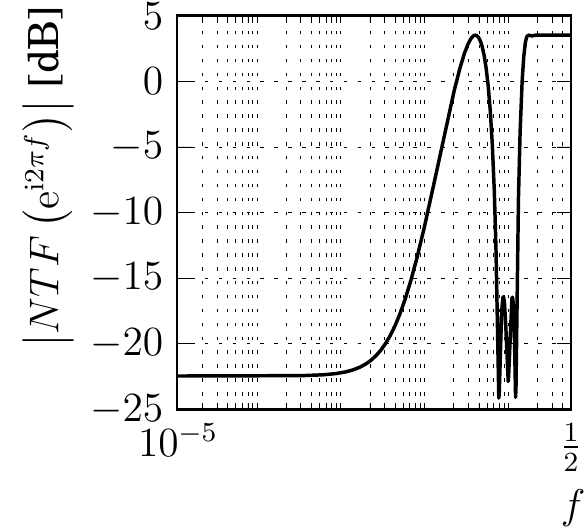}}\hfill    
  \end{tightcenter}
  \caption{Sample on-off specification for a multiband ΔΣ modulator with
    $\OSR=8$ \protect\subref{sfig:on-off-multi} and corresponding \ac{NTF}
    magnitude response \protect\subref{sfig:on-off-multi-ntf} obtained
    designing the \ac{NTF} with the proposed method for a 32\Us{nd} order
    \ac{FIR} \ac{NTF}.}
  \label{fig:on-off-multi}
\end{figure}

The proposed method makes the design of multiband modulators totally consistent
with that of single band ones. It is sufficient to take an on-off weighting
function that is non-null in multiple frequency intervals. As an example,
consider a two band modulator where the bands are $\unit[{[0,1]}]{kHz}$ and
$\unit[{[8,15]}]{kHz}$, so that the overall bandwidth is \unit[8]{kHz}. Assume
that the \ac{OSR} is 8, so that the sample frequency is \unit[128]{kHz}. The
on-off weighting function corresponding to this setup is $w(f)$ equal to $1$
for $f$ in $[0,\nicefrac{1}{64}] \cup [\nicefrac{1}{8},\nicefrac{15}{64}]$ and
null elsewhere, as in Fig.~\ref{sfig:on-off-multi}. The corresponding \ac{NTF}
obtained by the proposed design method for a 32\Us{th} order \ac{FIR} \ac{NTF}
is shown in Fig.~\ref{sfig:on-off-multi-ntf}.
 
Note that multiband signals cannot generally be managed by conventional design
flows. For instance, none of the \ac{NTF} design routines in the DELSIG toolbox
can straightforwardly deal with this case. Only very recent design approaches
such as \cite{Nagahara:TSP-60-6, Callegari:TCAS1-60-9} are suitable for this
task. For what concerns \cite{Callegari:TCAS1-60-9}, some comparison is
provided in the next Section~\ref{ssec:generalization}.

\subsection{Design of a modulator capable of optimally dealing with an
  imperfect filter for the removal of quantization noise}
\label{ssec:generalization}
\begin{figure}
  \begin{tightcenter}
    \includegraphics[width=\lw]{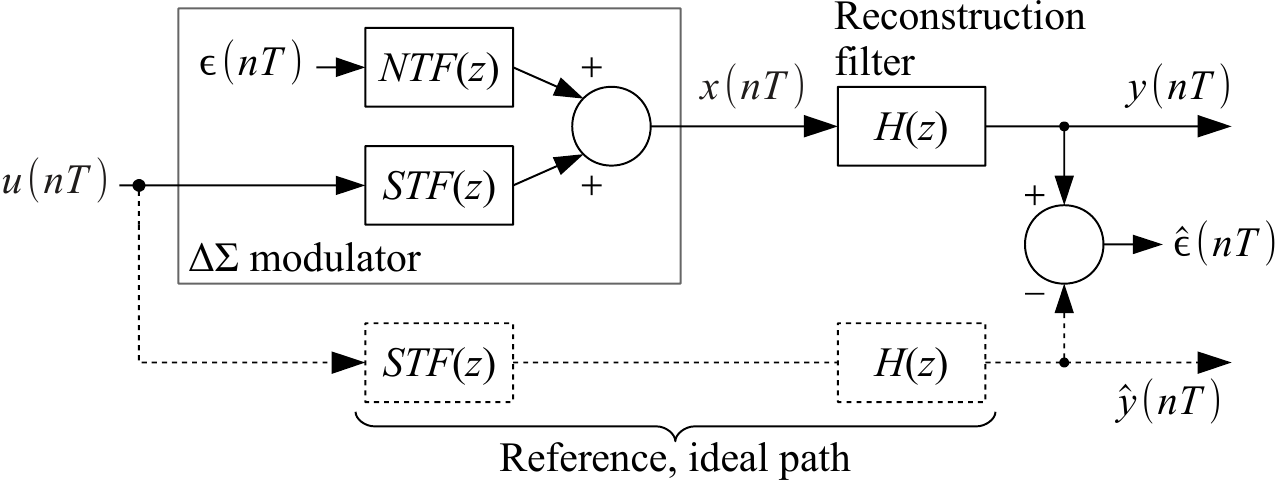}
  \end{tightcenter}
  \caption{Evaluation of the residual noise after the application of a
    reconstruction filter.}
  \label{fig:residual-noise}
\end{figure}

\begin{figure}
  \begin{tightcenter}\vskip -2ex%
    \setsubcapskip{-1ex}%
    \subfloat[\label{sfig:lp}]{%
      \includegraphics[scale=0.75]{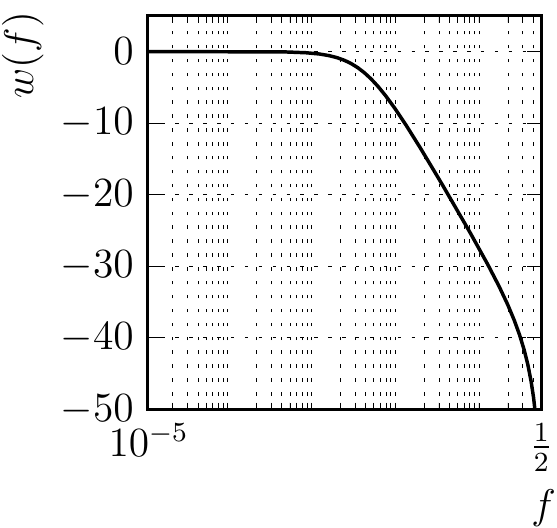}}\hfill
    \subfloat[\label{sfig:lp-ntf}]{%
      \includegraphics[scale=0.75]{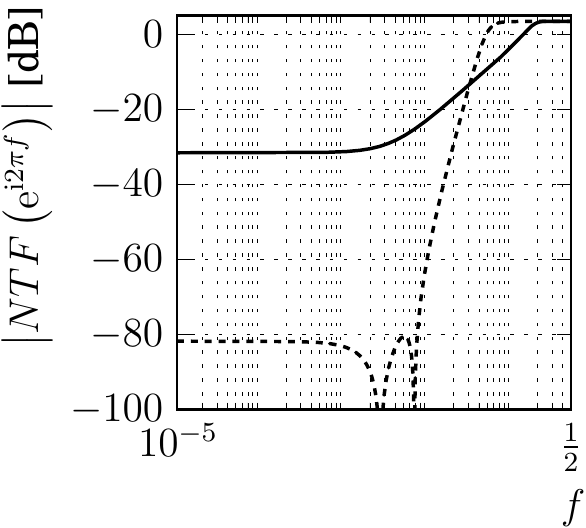}}\hfill    
  \end{tightcenter}
  \caption{Noise weighting function for an \ac{LP} ΔΣ modulator with $\OSR=64$
    and a 1\Us{st} order Butterworth reconstruction filter
    \protect\subref{sfig:lp} and corresponding optimal \ac{NTF} magnitude
    response \protect\subref{sfig:lp-ntf} for a 10\Us{th} order \ac{FIR}
    \ac{NTF} (solid). For comparison, a 4\Us{th} order conventional \ac{NTF}
    obtained with DELSIG is also shown (dashed).}
  \label{fig:lp}
\end{figure}

This case is the one tackled in \cite{Callegari:TCAS1-60-9}. It is reported
precisely to illustrate how the optimization procedures is ultimately the
same. Suppose that a \ac{DDSM} \ac{D/A} setup deals with signals defined in a
\unit[{[0,1]}]{kHz} band and that the reconstruction filter is a mere 1\Us{st}
order Butterworth unit with cut-off at \unit[1]{kHz}. Let the \ac{OSR} be
64. This relatively large value lets one approximate the reconstruction filter
in discrete time with good precision as $H(z)=12.12\times
10^{-3}\;\nicefrac{(z+1)}{(z-0.976)}$.  The residual quantization noise after
the reconstruction filter can be estimated as in Fig.~\ref{fig:residual-noise},
where a reference signal $\hat y(nT)$, obtained by filtering $u(nT)$ through
the \ac{STF} and $H(z)$ is subtracted from the actual output of the
reconstruction filter $y(nT)$ to give $\hat \epsilon(nT)$. Since the paths
through the \ac{STF} and $H(z)$ elide each other, the residual noise is
$\epsilon(nT)$ filtered through the \ac{NTF} and $H(z)$. Thus, its power is
\begin{equation}
  \sigma^2_{\hat \epsilon} = \int_0^{\frac{1}{2}} \Psi_{n}(f) 
  \abs{H\left(\ee^{\ii 2\pi f}\right)}^2\, df
  \label{eq:residual-noise}
\end{equation}
This is formally analogous to Eqn.~\eqref{eq:merit-factor} when
$w(f)=\abs{H\left(\ee^{\ii 2\pi f}\right)}^2$.  Unsurprisingly, the non-ideal
reconstruction filter determines a noise weighting equivalent to its magnitude
response. However, from a practical point of view, the computation of the
optimal modulator \ac{NTF} based on the \ac{IDTFT} operator is more efficient
and accurate than that proposed in \cite{Callegari:TCAS1-60-9}, the latter
being based on a computation practiced in the time domain from the filter
impulse response. For the sake of comparison, Fig.~\ref{sfig:lp-ntf} also shows
a reference \ac{NTF} obtained by DELSIG's \texttt{synthesizeNTF}. As in
\cite{Callegari:TCAS1-60-9}, the value of $\sigma^2_{\hat \epsilon}$ obtained
by the proposed method is almost \unit[5]{dB} better than the reference one.

\subsection{Design of a modulator with reduced dc noise}
\begin{figure}
  \begin{tightcenter}\vskip -2ex%
    \setsubcapskip{-1ex}%
    \subfloat[\label{sfig:on-off+dc}]{%
      \includegraphics[scale=0.74]{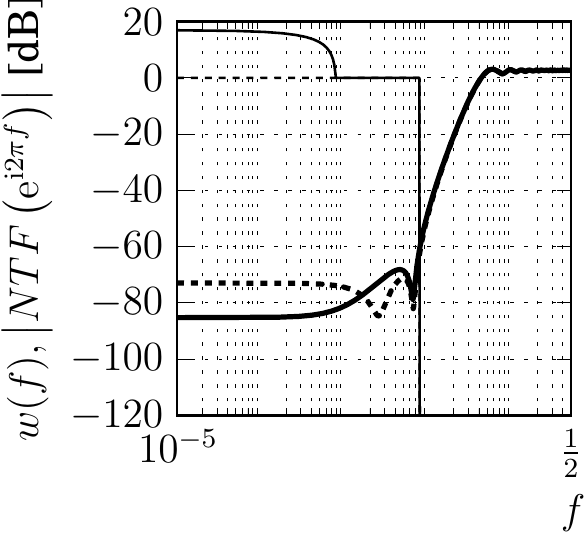}}\hfill
    \subfloat[\label{sfig:on-off+dc-pds}]{%
      \includegraphics[scale=0.74]{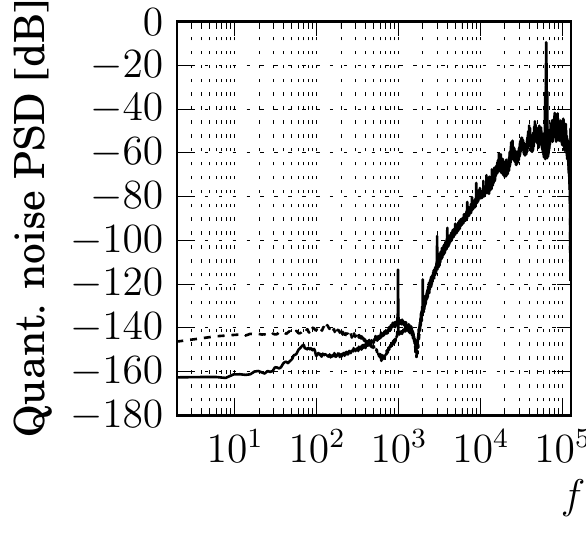}}\hfill    
  \end{tightcenter}
  \caption{In \protect\subref{sfig:on-off+dc}: noise weighting function (thin
    line) and \ac{NTF} magnitude response (thicker) for an \ac{LP} modulator
    capable of providing reduced noise close to dc (both solid). Data from the
    case in Sec.~\ref{ssec:on-off} provided for comparison (dashed). \ac{OSR}
    is 64 and the \ac{NTF} \ac{FIR} order is 32. In
    \protect\subref{sfig:on-off+dc-pds}: the corresponding \ac{PSD} of the
    shaped quantization noise, as obtained from time-domain simulations.}
  \label{fig:on-off+dc}
\end{figure}

In some testing or sensing applications, excitations are obtained by storing an
off-line generated ΔΣ sequence into a memory and playing it through a \ac{LP}
filter. Some of these applications are particularly sensitive to dc errors, so
that one may want a particularly low noise at low frequency values. The
proposed method makes it quite easy to achieve this result by starting with an
on-off noise weighting function (like in Fig.~\ref{sfig:on-off}) and then
raising it at low frequencies. Fig.~\ref{sfig:on-off+dc} shows how this can be
done together with the outcome in terms of \ac{NTF}.
Fig~\ref{sfig:on-off+dc-pds} shows that the method works, providing the actual
\ac{PSD} of the shaped quantization noise, as obtained from time-domain
simulations run on the nonlinear modulator model.

\subsection{Design of a psychoacoustically optimal coder for audio}
\label{ssec:psycho}
\acp{DDSM} can be used as coders in audio applications in view of highly
efficient switched-mode amplification \cite{Gaalaas:AD-40-6} or even to prepare
data for certain storage formats \cite{Reefman:CAES-2001}. In these
applications, one wants the in-band residual quantization noise to be minimally
audible. The way in which the ear reacts to sound pressure levels is extremely
complex since in human evolution the cochlea has learned to be most sensitive
to the sounds most important for survival. Such complexity is today captured by
\emph{psychoacoustic models}, commonly based on \emph{equal-loudness contours}
such as those in ISO 226 \cite{ISO226-2003}. ISO tables a family of functions
$L_p(f;L_N)$, obtained from experiments over a multitude of listeners and
parametrized over a \emph{perceived} sound loudness $L_N$. At each
characterized frequency, the actual acoustic power (sound pressure level)
necessary to get $L_N$ is returned. When a listener is exposed to a sound, the
overall perception can be quantified by integrating the sound \ac{PSD}
multiplied by a weighting profile obtained from $L_p(f;L_N)$ (approximately as
its inverse, although corrections accounting for nonlinear effects, limited
reach at high frequency and acoustic field features are commonly applied too
\cite{wannamaker:JAES-40-7}). Specific weightings have been developed for the
significant case where the sound is a spread-spectrum, low-level noise, a
notable one being the \emph{F-weighting}, based on $L_P(f;\unit[15]{phon})$,
and characterized as an analytic expression in \cite{wannamaker:JAES-40-7}. The
present design flow makes it easy to directly exploit weightings as the
\emph{F} one for the design of \acp{DDSM}.

Note that in principle also the magnitude response of every unit between the
coder and the hear (e.g. filters, amplifiers, loudspeakers), should be
considered. However, common weightings are already so steep at their edges
(below $\unit[20]{Hz}$ and above \unit[16]{kHz}), that unless pathological
components are used for these units, the influence of their band limitations on
noise perception is negligible.  Here, we use the \emph{F-weighting} as $w(f)$,
since this also favors comparison to previous results in the Literature.

\begin{figure}
  \begin{tightcenter}\vskip -2ex%
    \setsubcapskip{-1ex}%
    \subfloat[\label{sfig:psycho-ntf}]{%
      \includegraphics[scale=0.74]{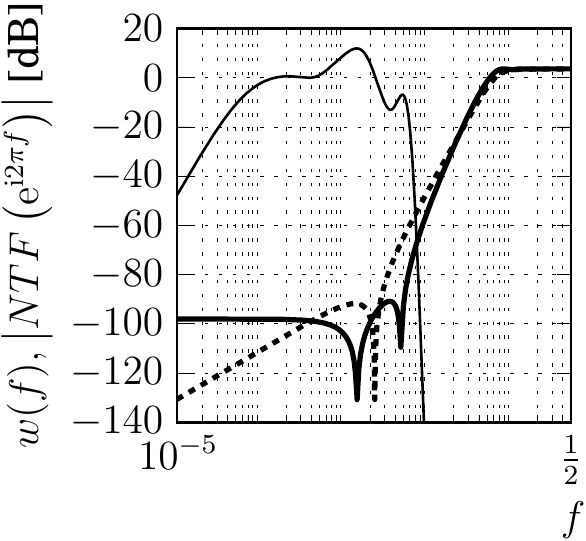}}\hfill
    \subfloat[\label{sfig:psycho-perceived}]{%
      \includegraphics[scale=0.74]{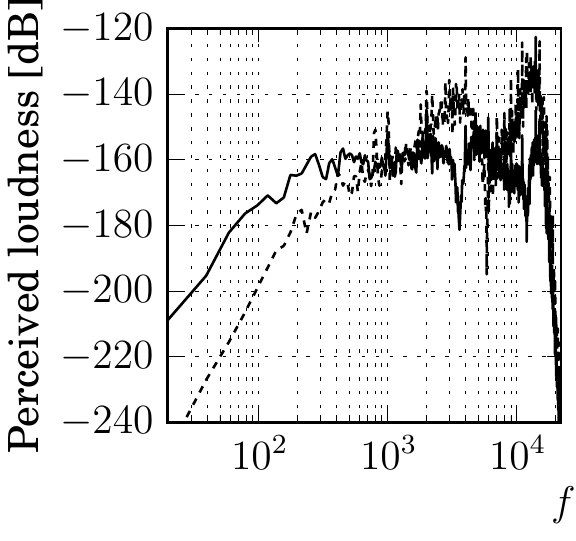}}   
  \end{tightcenter}
  \caption{In \protect\subref{sfig:psycho-ntf}: noise weighting function (thin
    line) and \ac{NTF} magnitude response (thicker) for a psychoacoustically
    optimal audio ΔΣ coder based on the \emph{F-Weighting} (30\Us{th}
    order). The \ac{NTF} from a 3\Us{rd} order \ac{DDSM} designed with the
    approach in \cite{Dunn:JAES-45-4} is provided for comparison
    (dashed). \ac{OSR} is 64. In \protect\subref{sfig:on-off+dc-pds}: the
    corresponding perceived loudness level of quantization noise as a function
    of frequency.}
  \label{fig:psycho}
\end{figure}

Fig~\ref{sfig:psycho-ntf} shows $w(f)$ together with a sample optimal \ac{NTF}
obtained with our design flow and another one obtained with the approach
described in \cite{Dunn:JAES-45-4} to design psychoacoustically optimal
\acp{DDSM}. Fig~\ref{sfig:psycho-perceived} shows the perceived noise spectra
obtained by time-domain simulations in the two cases. Clearly the \acp{NTF} are
quite different. This is due to many reasons.  The most evident one is that
\cite{Dunn:JAES-45-4} follows conventional design methods
\cite{Schreier:UDSDC-2004}, just plugging onto them a step where \ac{NTF} zeros
are adjusted \emph{on the unit circle} until the perceived noise \ac{PSD} is
made as flat as possible in the signal band. Consequently, the \ac{NTF} ends up
having a number of \emph{valleys} proportional to the modulator order. In our
approach, except for very low orders, the \ac{NTF} ends up always having as
many valleys as there are peaks in the weighting profile. Furthermore, because
our approach is less constrained, it succeeds better in lowering the \ac{NTF}
where this is important for the perceived SNR, namely where the weighting
profile (normalized to peak slightly above \unit[0]{dB}) and the \ac{NTF}
intersect.

\section{Conclusions}
\label{sec:conclusions}
A design flow for ΔΣ modulators has been illustrated, allowing quantization
noise to be shaped according to an arbitrary weighting profile. It is best
suited for \acp{DDSM} in applications where quantization noise (including the
residual in-band noise) is known to have a different impact at different
frequencies. An implementation is available at
\url{http://pydsm.googlecode.com} as part of an open source toolbox. This is
coded in Python/Scipy \cite{Oliphant:CS+E-9-3} and based on the CVXOPT/CVXPY
optimization core \cite{Andersen:OML-3-2011}. It is easy to install, can run on
Linux and Windows and only depends on free software. It can be used to replicate
the many examples in this paper or as a general design aid.



%

\bibliographystyle{SC-IEEEtran}
\bibliography{macros,IEEEabrv,various,sensors,analog,chaos}

\begin{thebibliography}{10}
\providecommand{\doi}[1]{DOI:#1}
\providecommand{\url}[1]{#1}
\csname url@samestyle\endcsname
\providecommand{\newblock}{\relax}
\providecommand{\bibinfo}[2]{#2}
\providecommand{\BIBentrySTDinterwordspacing}{\spaceskip=0pt\relax}
\providecommand{\BIBentryALTinterwordstretchfactor}{4}
\providecommand{\BIBentryALTinterwordspacing}{\spaceskip=\fontdimen2\font plus
\BIBentryALTinterwordstretchfactor\fontdimen3\font minus
  \fontdimen4\font\relax}
\providecommand{\BIBforeignlanguage}[2]{{%
\expandafter\ifx\csname l@#1\endcsname\relax
\typeout{** WARNING: IEEEtran.bst: No hyphenation pattern has been}%
\typeout{** loaded for the language `#1'. Using the pattern for}%
\typeout{** the default language instead.}%
\else
\language=\csname l@#1\endcsname
\fi
#2}}
\providecommand{\BIBdecl}{\relax}
\BIBdecl

\bibitem{Harris:WET-2003}
F.~Harris, ``Sigma-delta converters in communication systems,'' in \emph{Wiley
  Encyclopedia of Telecommunications}, J.~G. Proakis, Ed.\hskip 1em plus 0.5em
  minus 0.4em\relax John Wiley \& Sons, Inc., 2003, vol.~IV, pp. 2227--2247.

\bibitem{Pamarti:TCAS1-54-3}
S.~Pamarti, J.~Welz, and I.~Galton, ``Statistics of the quantization noise in
  1-bit dithered single-quantizer digital delta-sigma modulators,''
  \emph{{IEEE} Trans. Circuits Syst. {I}}, vol.~54, no.~3, pp. 492--503, Mar.
  2006.

\bibitem{Su:TCAS2-56-12}
P.-E. Su and S.~Pamarti, ``Fractional-n phase-locked-loop-based frequency
  synthesis: A tutorial,'' \emph{{IEEE} Trans. Circuits Syst. {II}}, vol.~56,
  no.~12, pp. 881--885, Dec. 2009.

\bibitem{Norsworthy:DSDC-1996}
S.~R. Norsworthy, R.~Schreier, and G.~C. Temes, Eds., \emph{Delta-Sigma Data
  Converters: Theory, Design, and Simulation}.\hskip 1em plus 0.5em minus
  0.4em\relax Wiley-IEEE Press, 1996.

\bibitem{Schreier:UDSDC-2004}
R.~Schreier and G.~C. Temes, \emph{Understanding Delta-Sigma Data
  Converters}.\hskip 1em plus 0.5em minus 0.4em\relax Wiley-IEEE Press, 2004.

\bibitem{Callegari:TCAS1-60-9}
S.~Callegari and F.~Bizzarri, ``Output filter aware optimization of the noise
  shaping properties of ΔΣ modulators via semi-definite programming,''
  \emph{{IEEE} Trans. Circuits Syst. {I}}, vol.~60, no.~9, pp. 2352--2365, Sep.
  2013.

\bibitem{Kenney:AICSP-3}
J.~G. Kenney and L.~R. Carley, ``Design of multibit noise-shaping data
  converters,'' \emph{Analog Integrated Circuits and Signal Processing},
  vol.~3, pp. 259--272, 1993.

\bibitem{Schreier:DELSIG}
\BIBentryALTinterwordspacing
R.~Schreier, \emph{The Delta-Sigma Toolbox}, Analog Devices, 2011, release 7.4,
  also known as ``{DELSIG}''. [Online]. Available:
  \url{http://www.mathworks.com/matlabcentral/fileexchange/}
\BIBentrySTDinterwordspacing

\bibitem{Callegari:TSP-58-12}
S.~Callegari, F.~Bizzarri, R.~Rovatti, and G.~Setti, ``On the approximate
  solution of a class of large discrete quadratic programming problems by
  {$\Delta\Sigma$} modulation: the case of circulant quadratic forms,''
  \emph{{IEEE} Trans. Signal Process.}, vol.~58, no.~12, pp. 6126--6139, Dec.
  2010.

\bibitem{Bizzarri:ISCAS-2010}
F.~Bizzarri and S.~Callegari, ``A heuristic solution to the optimisation of
  flutter control in compression systems (and to some more binary quadratic
  programming problems) via {$\Delta\Sigma$} modulation circuits,'' in
  \emph{Proc.\@ of ISCAS'10}, Paris, FR, May 2010.

\bibitem{Callegari:ICECS-2012}
S.~Callegari and F.~Bizzarri, ``Should ΔΣ modulators used in ac motor drives
  be adapted to the mechanical load of the motor?'' in \emph{Proceedings of
  IEEE ICECS 2012}, Seville (ES), Dec. 2012, pp. 849--852.

\bibitem{Bizzarri:ISCAS-2012}
F.~Bizzarri, S.~Callegari, and G.~Gruosso, ``Towards a nearly optimal synthesis
  of power bridge commands in the driving of {AC} motors,'' in
  \emph{Proceedings of ISCAS 2012}, Seoul, May 2012, pp. 2119--2122.

\bibitem{ISO226-2003}
\emph{Acoustics — Normal equal-loudness-level contours}, ISO Std. 226,
  Rev.~2, Aug. 2003.

\bibitem{Lee:Thesis-1987}
W.~L. Lee, ``A novel high order interpolative modulator topology for high
  resolution oversampling {A/D} converters,'' Master's thesis, Massachussets
  Institute of Technology, 1987.

\bibitem{Dunn:JAES-45-4}
C.~Dunn and M.~Sandler, ``Psychoacoustically optimal sigma delta modulation,''
  \emph{Journal of the Audio Engineering Society (AES)}, vol.~45, no.~4, pp.
  212--223, Apr. 1997.

\bibitem{Nagahara:TSP-60-6}
M.~Nagahara and Y.~Yamamoto, ``Frequency domain {M}in-{M}ax optimization of
  noise-shaping {D}elta-{S}igma modulators,'' \emph{{IEEE} Trans. Signal
  Process.}, vol.~60, no.~6, pp. 2828--2839, Jun. 2012.

\bibitem{wannamaker:JAES-40-7}
R.~A. Wannamaker, ``Psychoacoustically optimal noise shaping,'' \emph{Journal
  of the Audio Engineering Society}, vol.~40, no. 7/8, p. 611–620, 1992.

\bibitem{Oppenheim:DTSP-1989}
A.~V. Oppenheim and R.~W. Schafer, \emph{Discrete-Time Signal
  Processing}.\hskip 1em plus 0.5em minus 0.4em\relax Prentice-Hall, 1989.

\bibitem{Iwasaki:TAC-50-1}
T.~Iwasaki and S.~Hara, ``Generalized {KYP} lemma: Unified frequency domain
  inequalities with design applications,'' \emph{{IEEE} Trans. Autom. Control},
  vol.~50, no.~1, pp. 41--59, Jan. 2005.

\bibitem{Boyd:LMISCT-1994}
S.~Boyd, L.~El~Ghaoui, E.~Feron, and V.~Balakrishnan, \emph{Linear Matrix
  Inequalities in System and Control Theory}, ser. SIAM studies in applied
  mathematics.\hskip 1em plus 0.5em minus 0.4em\relax Philadelphia: SIAM, 1994,
  vol.~15.

\bibitem{Andersen:OML-3-2011}
\BIBentryALTinterwordspacing
M.~Andersen, J.~Dahl, Z.~Liu, and L.~Vandenberghe, ``Interior-point methods for
  large-scale cone programming,'' in \emph{Optimization for Machine Learning},
  ser. Neural Information Processing series, S.~Sra, S.~Nowozin, and S.~J.
  Wright, Eds.\hskip 1em plus 0.5em minus 0.4em\relax MIT Press, Sep. 2011,
  ch.~3, pp. 55--84. [Online]. Available:
  \url{http://www.ee.ucla.edu/~vandenbe/publications/mlbook.pdf}
\BIBentrySTDinterwordspacing

\bibitem{Gaalaas:AD-40-6}
E.~Gaalaas, ``Class {D} audio amplifiers: What, why, and how,'' \emph{Analog
  Dialogue}, vol.~40, no.~6, pp. 1--7, Jun. 2006.

\bibitem{Reefman:CAES-2001}
D.~Reefman and P.~Nuijten, ``Why {D}irect {S}tream {D}igital is the best choice
  as a digital audio format,'' in \emph{Proc.\@ of the 110th Convention of the
  Audio Engineering Society}, 2001.

\bibitem{Oliphant:CS+E-9-3}
T.~E. Oliphant, ``Python for scientific computing,'' \emph{{IEEE} Comput. Sci.
  Eng.}, vol.~9, no.~3, pp. 10--20, May 2007.

\end{thebibliography}

\end{document}